\documentstyle[aps]{revtex}
\input psfig.tex

\begin{document}
\title{Turbulent Pair Diffusion}
\draft
\author{F. Nicolleau}
\address{
The University of Sheffield,
Department of Mechanical Engineering,
Mappin Street, Sheffield, S1 3JD, UK
}
\author{J. C. Vassilicos}
\address{
Imperial College of Science, Technology and Medicine,
Department of Aeronautics,
Prince Consort Road, South Kensington,
London, SW7 2BY, UK
}
\date{\today}



\maketitle

\begin{abstract}

Kinematic Simulations of turbulent pair diffusion in planar turbulence
with a $k^{-5/3}$ energy spectrum reproduce the results of the
laboratory measurements of Jullien {\it et al} Phys. Rev. Lett. {\bf
82}, 2872 (1999), in particular the stretched exponential form of the
PDF of pair separations and their correlation functions. The root mean
square separation is found to be strongly dependent on initial
conditions for very long stretches of times. This dependence is
consistent with the topological picture of turbulent pair diffusion
where pairs initially close enough travel together for long stretches
of time and separate violently when they meet straining regions around
hyperbolic points. A new argument based on the divergence of accelerations
is given to support this picture. 
 
\end{abstract}


The rate with which pairs of points separate in phase space or in
physical space is of central importance to the study of dynamical
systems. Pairs of points in the phase space of a low-dimensional
chaotic dynamical system separate exponentially. This is the
celebrated butterfly effect: dynamics are extremely sensitive to
initial conditions. Pairs of fluid elements in fully choatic flows
\cite{ottino} also separate exponentially. However, in fully developed
homogeneous and isotropic turbulence, Richardson's law
\cite{richardson} stipulates that fluid element pairs separate on
average algebraically and in such a way that their separation
statistics in a certain range of times are the same irrespective of
initial conditions. Richardson's law is therefore a remarkable claim
of universality. Specifically, it stipulates that in a range of times
where the root mean square separation $\overline{ \Delta^{2}} ^{1/2}$
is larger than the Kolmogorov length-scale $\eta$ and smaller than the
integral length-scale $L$, $\overline{ \Delta^{2}}$ is increasingly
well approximated by

\begin{equation}
\overline{ \Delta^{2}} = G_{\Delta}\epsilon t^{3}
\label{eq:1}
\end{equation}
for increasing values of $L/\eta$, where $t$ is time, $\epsilon$ is the 
kinetic energy rate of dissipation per unit mass and $G_{\Delta}$ is a 
universal dimensionless constant. 

Richardson accompanied his empirical law (\ref{eq:1}) with a
prediction for the probability density function (PDF) of pair
separations $\Delta$. The effective diffusivity approach leading to
this prediction was criticised by Batchelor \cite{batchelor} who
developed a different approach leading to (\ref{eq:1}) but also to a
different form of the PDF.  Kraichnan \cite{kraichnan} derived yet
another expression for the PDF based on his Lagrangian history direct
interaction approximation and so did Shlesinger {\it et al}
\cite{shlesinger} on the assumption that turbulent pair diffusion is
well described by L\'evy walks.

Setting $\sigma (t) \equiv \overline{\Delta^{2}} ^{1/2}$ and $r\equiv
\Delta/\sigma$, the PDFs of $\Delta$ predicted by Richardson
\cite{richardson}, Batchelor \cite{batchelor} and Kraichnan
\cite{kraichnan} are all of the form

\begin{equation}
P(\Delta, t)\sim \sigma^{-1}\exp (-\alpha r^{\beta})
\label{eq:2}
\end{equation}
with different values of the dimensionless parameters $\alpha$ and
$\beta$. Richardson's prediction for the exponent $\beta$ is
$\beta=2/3$, Batchelor's is $\beta=2$ and Kraichnan's is $\beta=4/3$.
The Lagrangian modelling approach of Shlesinger {\it et al}
\cite{shlesinger} leads to a totally different, in fact algebraic, PDF
form. More recently Jullien {\it et al} \cite{tabeling} reported
laboratory measurements of $P(\Delta, t)$ which are well fitted by
(\ref{eq:2}) with $\alpha \approx 2.6$ and $\beta = 0.5 \pm
0.1$. These laboratory measurements invalidate the PDF predictions of
Batchelor, Kraichnan and Shlesinger {\it et al.} and might raise a
question mark over the PDF prediction of Richardson even though they
can be considered consistent with it if we account for experimental
uncertainties. Jullien {\it et al} \cite{tabeling} also observed that
fluid element pairs stay close to each other for a long time until
they separate quite suddenly, a behaviour which seems qualitatively at
odds with the effective diffusivity approach adopted by Richardson
\cite{richardson} to derive (\ref{eq:2}) with $\beta=2/3$. In
particular, they measured the Lagrangian autocorrelation function of
pair separations $R(t,\tau) \equiv <\Delta (t) \Delta (t+\tau)>$ for
$-t\le \tau \le 0$ and found a Lagrangian pair correlation time
$\tau_c \approx 0.6t$ which is surprisingly long. In this paper we
report that Kinematic Simulation (KS) \cite{KS} reproduces the
experimental results of Jullien {\it et al} \cite{tabeling}. KS is a
Lagrangian model of turbulent diffusion which is distinct from L\'evy
walks \cite{shlesinger} and makes no use of Markovianity assumptions
so that it cannot be reduced to an effective diffusivity approach such
as Richardson's \cite{richardson}. Furthermore, the observation that
fluid element pairs travel close together for long stretches of time
until they separate quite suddenly has in fact already been made using
KS \cite{KS}.

KS Lagrangian modelling consists in integrating fluid element
trajectories by solving $\frac{d {\bf x}(t)}{dt}={\bf u}({\bf x}(t),
t)$ in synthesised velocity fields ${\bf u}({\bf x},
t)$. Statistically homegeneous, isotropic and stationary KS velocity
fields are superpositions of random Fourier modes \cite{KS}. KS
velocity fields are gaussian but {\it not} delta-correlated in time
\cite{KS}, and this non-Markovianity is an essential ingredient in KS.
The Lagrangian measurements of \cite{tabeling} were made in an inverse
cascade two-dimensional turbulent flow. Our KS velocity field is
therefore prescribed to be planar and given by

\begin{eqnarray}
{\bf u} =\sum_{m=1}^{m=M}
[{\bf A}_m \times {\hat{\bf k}_m}\cos{({\bf k}_m\cdot{\bf x}+\omega_m t)}
\nonumber\\
+{\bf B}_m \times {\hat{\bf k}_m}\sin{({\bf k}_m\cdot{\bf x}+\omega_m t)}]
\label{eq:3}
\end{eqnarray}
where $M=500$ is the number of modes, ${\hat{\bf k}_m}$ is a random
unit vector (${\bf k}_m=k_m{\hat{\bf k}_m}$) normal to the plane of the flow
whilst the vectors ${\bf A}_m$ and ${\bf B}_m$ are in that
plane.  The random choice of directions for the $m^{th}$ wavemode is
independent of the choices of the other wavemodes.  Note that the
velocity field ${\bf u}$ is incompressible by construction. The
amplitudes $A_m$ and $B_m$ of the vectors ${\bf A}_m$ and ${\bf B}_m$
are determined by the Kolmogorov energy spectrum $E (k)$ via the
relations $A_m^2=B_m^2=E (k_m) \Delta k_m$ where $\Delta
k_m=(k_{m+1}-k_{m-1})/2$. Finally the unsteadiness
frequencies $\omega_m$ are determined by the eddy turnover time of
wavemode $m$, that is $\omega_m=0.5\sqrt{k_m^3 E(k_m)}$.

The Lagrangian measurements in \cite{tabeling} were made when the
two-dimensional flow had developed an inverse cascade with a
well-defined $k^{-5/3}$ energy spectrum. The energy spectrum we have
therefore chosen for this study is $E (k)\approx
\frac{2u'^{2}}{3L^{2/3}}k^{-5/3}$ in the range $\frac{2\pi}{L}\le k
\le \frac{2\pi}{\eta}$ and equal to 0 outside this range ($u'^{2}$ is
the total kinetic energy of the turbulence). The $M=500$ wavenumbers
are algebraically distributed between $\frac{2\pi}{L}$ and
$\frac{2\pi}{\eta}$.
The eddy turnover time at the largest wavenumber $\frac{2\pi}{\eta}$ can be
considered to correspond to a Kolmogorov time scale $\tau_{\eta}$.

\begin{figure}[h]
\psfig{file=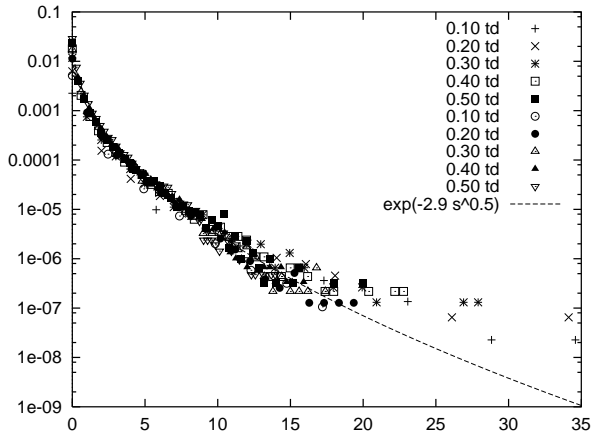 ,height=6cm,clip=}
\caption{Semi-log plot of $\sigma p(r)$ as a function of $r = {\Delta \over \sigma}$ 
in the case ${L \over \eta} = 1691$.
${\Delta_0 \over \eta} = 0.1$ and ${t u'\over L}=$
$+$ 0.10, $\times$ 0.20, $\ast$ 0.30, empty box 0.40, black box 0.5.
${\Delta_0 \over \eta} = 1$ and ${tu'\over L}=$
$\odot$ 0.10,
$\bullet$ 0.20,
$\bigtriangleup$ 0.30,
black triangle 0.4 and
$\bigtriangledown$ 0.5.
The solid line is $\sigma p(r) \sim e^{-2.9 r^{0.5}}$.}
\label{figure1}
\end{figure}

The inertial range ratio $\frac{L}{\eta}$ is $O(10)$ in the laboratory
experiment of \cite{tabeling} but here we have also run simulations
with $\frac{L}{\eta} = 10, 100, 1691, 11180, 38748, 250000$. For
initial pair separations $\Delta_0$ smaller or equal to $\eta$ our KS
integrations lead to $\sigma P(\Delta, t)\sim \exp (-\alpha
r^{\beta})$ where $r=\Delta /\sigma (t)$ with $2.6 \le \alpha \le 3$
and $0.46 \le \beta \le 0.5$ in very good agreement with the
laboratory results of \cite{tabeling} and for all the $\frac{L}{\eta}$
values that we tried (see example in Figure 1). (We record, however,
that this PDF does seem to depend on the initial separation $\Delta_0$
when $\Delta_0 > \eta$.) It has already been noted in \cite{fv00} that
KS gives non-gaussian stretched exponential PDFs of pair separations
without, however, estimating $\alpha$ and $\beta$. The synthetic
velocity fields of \cite{romans} lead to the Richardson stretched
exponential form with $\beta = 2/3$. An approach based on asymmetric
Levy walks \cite{sokolov} gives rise to stretched exponential forms of
$\sigma P(\Delta, t)$ where $\beta$ can be tuned as a function of a
persistence parameter.
\begin{figure}[h]
\psfig{file=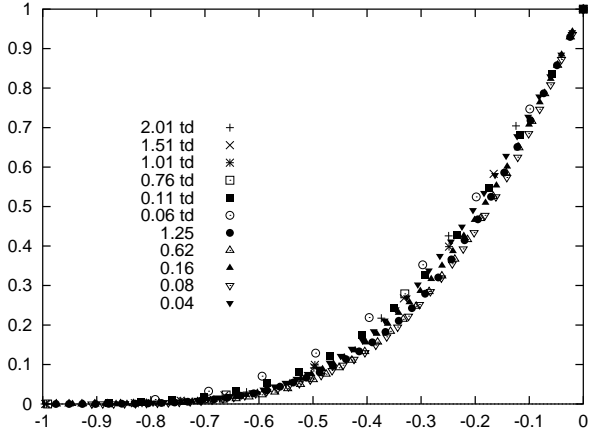,height=6cm,clip=}
\caption{Lagrangian separation correlation factor $R(t,\tau)/ \sigma^2(t)$
as a function of ${\tau \over t}$ in the case ${L \over \eta} = 1691$. 
${\Delta_0 \over \eta} = 1$ and
${t u'\over L}=$
$+$ 2.01,
$\times$ 1.51,
$\ast$ 1.01,
empty box 0.76,
black box 0.11,
$\odot$ 0.06.
${\Delta_0 \over \eta} = 0.1$ and ${t u'\over L}=$
$\bullet$ 1.25,
$\bigtriangleup$ 0.62,
black triangle 0.16,
$\bigtriangledown$ 0.08,
black triangle down 0.04.
}
\label{figure2}
\end{figure}

Following \cite{tabeling} we also calculate correlation functions of
pair separations, i.e. $R(t,\tau) \equiv <\Delta (t) \Delta (t+
\tau)>$ for $-t\le \tau \le 0$ and with $\Delta_0$ equal to $\eta$ in
one set of runs and $0.1\eta$ in another (the choice of $\Delta_0$ in
\cite{tabeling} is within this range). The laboratory results
\cite{tabeling} show that $R(t,\tau)/\sigma^{2}(t)$ is a function of
$\tau / t$ and exactly the same collapse is found here with KS (Figure
2). We calculate a Lagrangian correlation time from $R(t,\tau)$ in the
way done in \cite{tabeling} and we obtain $\tau_c \approx 0.45t$ from
Figure 2. This value $0.45$ is indeed the constant asymptotic value
that we obtain for all large enough scale ratios $\frac{L}{\eta}$, i.e. 
$\frac{L}{\eta}\ge O(10)$, and it compares sufficiently well with
$\tau_c \approx 0.6t$ in the laboratory experiment \cite{tabeling}. 
The agreement is therefore good and KS leads to
the same conclusion, effectively  as the laboratory experiment, 
that pair separations remember about half their history.
\begin{figure}[h]
\psfig{file=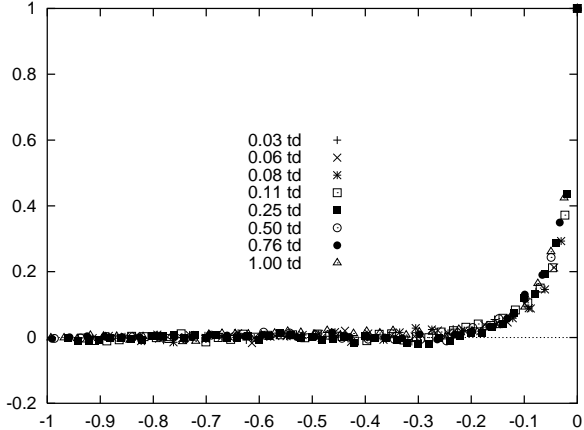,height=6cm,clip=}
\caption{Non dimensional diagonal Lagrangian velocity correlation 
$D_{11}(t,\tau)/D_{11}(t,0)$ as a function of $\tau / t$. Same case
as Figure~\ref{figure2}.
${t u'\over L}=$
$+$ 0.03, $\times$ 0.06, $\ast$ 0.08, empty box 0.11, black box 0.25,
$\odot$ 0.50, $\bullet$ 0.76, $\bigtriangleup$ 1.}
\label{figure3}
\end{figure}

The last set of statistics measured by \cite{tabeling} are Lagrangian
correlations of pair velocity differences, i.e. $D_{ij} \equiv
<V^{L}_i (t)V^{L}_j (t+\tau)>$ with $-t\le \tau \le 0$, where $V^{L}_i
(t)$ denotes the $i$th component of the Lagrangian relative velocity
between a pair of fluid elements. We calculate these same statistics
using our KS model and find that $D_{ij}$ remains close to 0 for $i
\not = j$, that $D_{11}(t,\tau)/D_{11}(t,0)$ and
$D_{22}(t,\tau)/D_{22}(t,0)$ are functions of $\tau/t$ (see Figure 3) and that this
collapse is the same for $D_{11}$ and $D_{22}$ again
in agreement with the laboratory results of \cite{tabeling}.

\begin{figure}
\psfig{file=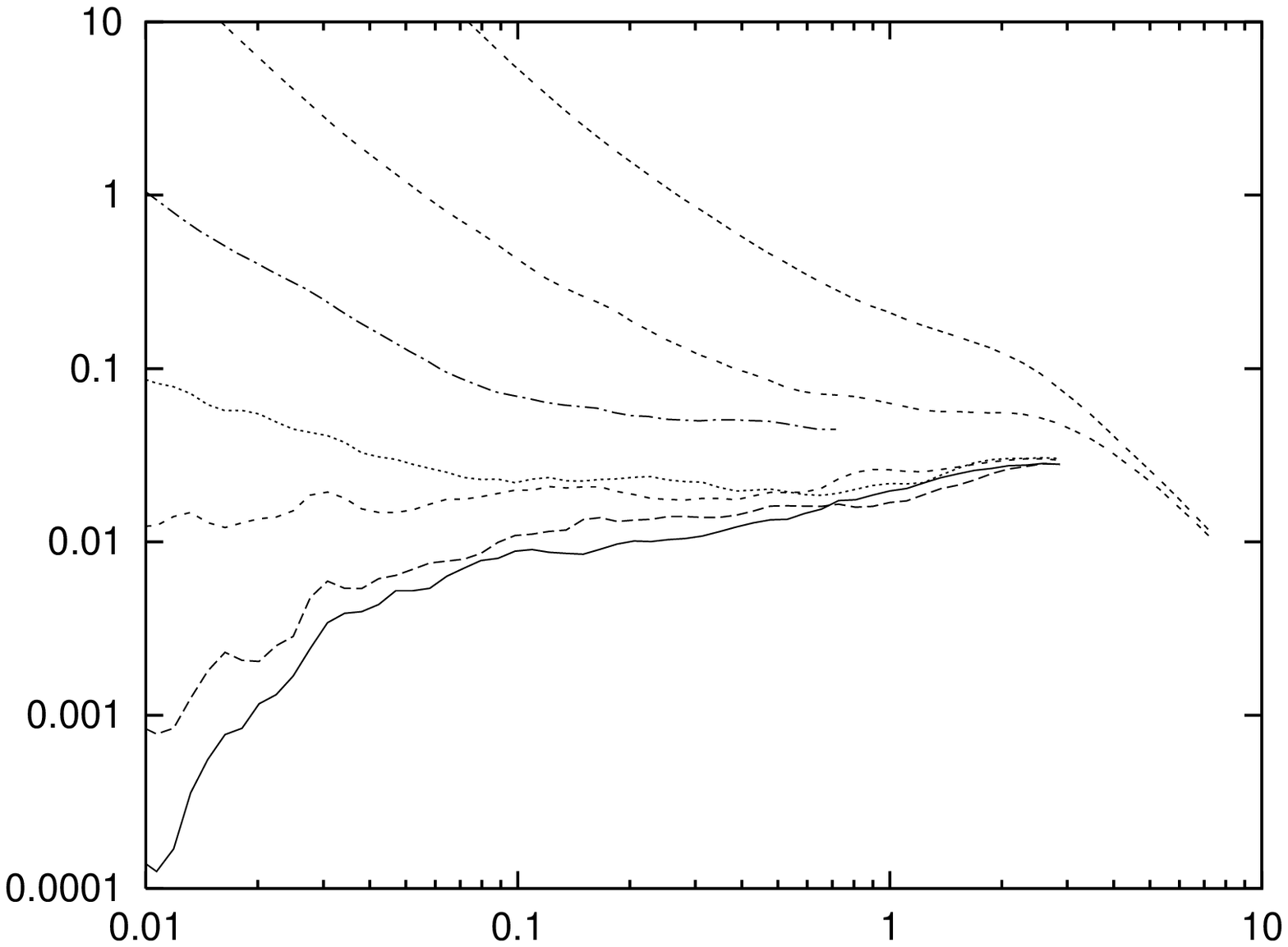,height=6cm,clip=}
\vspace*{-6cm}
\hspace*{8cm}
\psfig{file=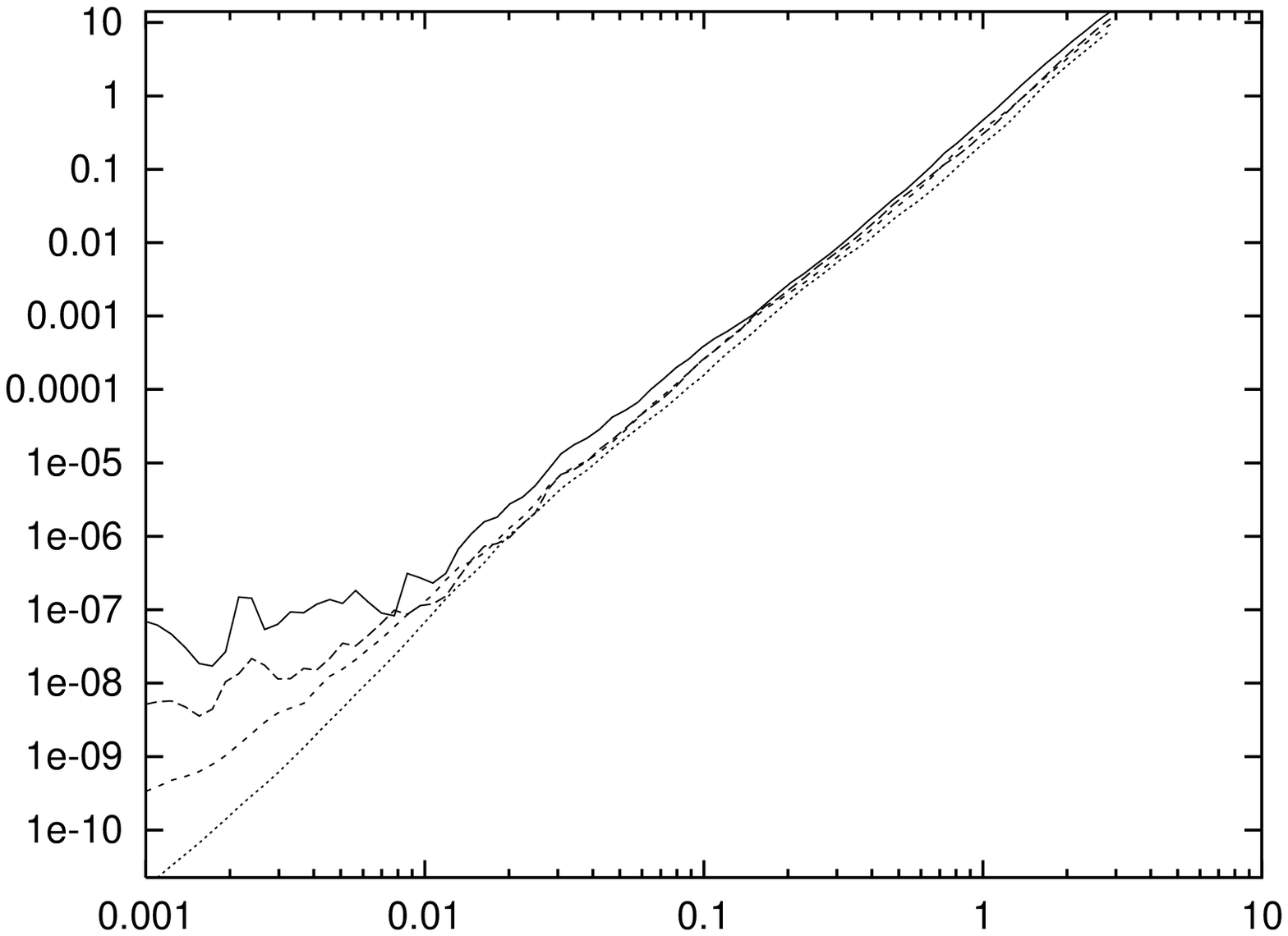,height=6cm,clip=}
\caption{Pair diffusion as a funtion of time for ${L \over \eta} = 38748$ 
and $\tau_{\eta} = 0.0027 {L \over u'}$
and
different initial separations.
a) ${\overline{(\Delta - \Delta_0)^{2}} \over u'^3 t^3 / L}$ as a function of 
$t {u' \over L}$, 
from top to botom 
${\Delta_0 \over \eta} = 1000$, 100, 10, 1, 0.1, 0.01 and 0.001.
b) ${\overline{(\Delta - \Delta_0)^{2}} \over f }$ as a function of $t {u' \over L}$ 
for ${\Delta_0 \over \eta} = 1$, 0.1, 0.01 and 0.001.}
\label{figure4}
\end{figure}

Having validated our KS Lagrangian model of pair diffusion in planar
turbulence against the laboratory experiment of \cite{tabeling}, we
now turn our attention to Richardson's law (1) and the claim of
universality that it is based on. We do indeed observe this law over
the entire inertial range of times $\tau_{\eta} < t < L/u'$, 
but only for initial separations $\Delta_0$
between $\eta$ and $0.1\eta$, and this for all the ratios
$\frac{L}{\eta}$ that we tried (see Figure 4a). 
Of course this ratio should be large enough, otherwise
Richardson's law is not observed for any $\Delta_0$, but it is
surprising that Richardson's law is so $\Delta_0$-specific even at
enormous values of $\frac{L}{\eta}$ reaching $ O (10^5) $.

When $\Delta_0 \le  \eta$, Richardson's law (1) is observed over the
limited large scale range $0.2 \frac{L}{u'}$ to $\frac{L}{u'}$
(Figure 4), and the coefficient 0.2 seems to have no dependence on
$\frac{L}{\eta}$ in our simulations as long as $L/ \eta$ is order $10^3$ or larger, 
so if it has one it must be
weak. In the remainder of the inertial range between $\tau_{\eta}$ and
$0.2 \frac{L}{u'}$,  the time dependencies of
$\overline{(\Delta - \Delta_0 )^{2}}$ and $\overline{\Delta^{2}}$ are
different from Richardson's (1) and different for different values of
$\Delta_0 \le \eta$ (Figure 4), even at extremely high
$\frac{L}{\eta}$. We tried to replace $t$ by $t-t_0 (\Delta_0 )$,
where $t_0 (\Delta_0 )$ is a virtual origin significantly smaller than
$0.2 \frac{L}{u'}$ but did not recover Richardson's law (1),
particularly since the discrepancies we observe are over such wide
time ranges. When $\Delta_0$ is significantly larger than $\eta$ there
is no clear indication of a Richardson law at all (Figure 4).

We have carefully studied the time dependence of $\overline{(\Delta -
\Delta_0 )^{2}}$ in the range between $\tau_{\eta}$ to $\frac{L}{u'}$
for different values of $\Delta_0 \le \eta$ and have found the following
formula to collapse the data in that range (see Figure 4b):
\begin{equation}
\overline{ (\Delta -\Delta_0 )^{2}} = G_{\Delta} 
\frac{u'^{3}}{L}t^{3} f(t, \Delta_0 )
\label{eq:4}
\end{equation}
where the dimensionless function $f$ is given by 
(using $T\equiv 0.2 \frac{L}{u'}$)
\begin{equation}
f(t, \Delta_0 )= \exp \left [ \frac{\ln [\frac{5}{G_{\Delta}}
( \Delta_0 / \eta)^2]}{2\ln (\tau_{\eta}/ T)} \left ( \ln(t/ T) - 
\sqrt{\ln^{2}(t/ T)+2} \right ) \right ].
\label{eq:5}
\end{equation}
Note that $f(t, \Delta_0 )$ tends to 1 when $t$ is between $T$ and
$\frac{L}{u'}$ and $L/\eta \to \infty$ (i.e. $T/\tau_{\eta} \to
\infty$).  The Richardson constant $G_{\Delta}$ is determined from the
value of $\frac{\overline{ (\Delta -\Delta_0 )^{2}}}{(u'^{3}/L)t^3}$
in the range $0.2 \frac{L}{u'}$ to $\frac{L}{u'}$ and we find
$G_{\Delta} \approx 0.03$ for large enough scale ratio
$\frac{L}{\eta}$ (of order $10^3$ and larger). We should stress that
in KS, $G_{\Delta}$ effectively contains both the original Richardson
constant as in $G_{\Delta}\epsilon t^3$ but also the constant of
proportionality relating the kinetic energy dissipation rate to
$\frac{u'^{3}}{L}$. We therefore retain the orders of magnitude of
$G_{\Delta}$ obtained by KS but not the actual values.

Integrations of $\overline{ \Delta^{2}}$ in Direct Numerical
Simulations (DNS) of two-dimen\-sional turbulence in the inverse energy
cascade regime also show a strong dependence on $\Delta_0$, even when
the simulations are very highly resolved \cite{boffetta}. Nowadays,
such DNS cannot reach well-defined $-5/3$ ranges over more
than two decades, i.e.  $L/\eta = O(100)$, and this at the very
highest resolutions currently available. The deviations from
Richardson's law (1) observed when $L/\eta = O(100)$ might perhaps be
due to edge effects ($L/\eta$ too small to reach the asymptotic
Richardson's law expected to be valid for $L/\eta \gg 1$). But can
this also be the case in our KS where $L/\eta$ reaches $O(10^5)$?
Clearly one cannot answer this question with numerical simulations
except if in the future KS and/or DNS runs with even higher
$L/\eta$ eventually converge to Richardson's law without $\Delta_0$ -
dependencies.

Nevertheless, the success of our KS to reproduce the laboratory
observations of \cite{tabeling} and its failure to retrieve
Richardson's law without $\Delta_0$ - dependencies even at extremely
high $L/\eta$ does raise the question of the validity of Richardson's
universality and of the locality assumption that it is based on
\cite{KS}, even asymptotically for arbitrarily high $L/\eta$. In
general, $\overline{ \Delta^{2}}$ is a function of $t$, $L$, $\eta$,
$\Delta_0$ and $u'$ in KS, and the Richardson locality assumption
adapted for KS states that, for large enough $L/\eta$, ${d\over
dt}\overline{ \Delta^{2}}$ should only depend on $\overline{
\Delta^{2}}$ and $E(k)$ at $k=2\pi / \sqrt{\overline{ \Delta^{2}}}$
when $max(\eta , \Delta_0 ) \ll \sqrt{\overline{ \Delta^{2}}} \ll
L$. Fung \& Vassilicos (1998) \cite{KS} found this assumption to be
valid in planar KS for different spectral exponents $p$ between 1 and
2 ($E(k)\sim k^{-p}$) but specifically for $\Delta_0 = \eta /2$ and
unsteadiness parameter $\lambda = O(1)$ and smaller than 1 in
$\omega_m=\lambda \sqrt{k_m^3 E_n(k_m)}$. The direct consequence of
this assumption is that $\overline{ \Delta^{2}} \sim t^{\gamma}$ with
$\gamma = {4\over 3-p}$ which is indeed observed in KS for different
values of $p$ but only for $\Delta_0$ close to and below $\eta$
\cite{KS}. What could invalidate locality and Richardson's law for
$\Delta_0$ very different from $\eta$?

The low values of $G_{\Delta}$ and the very large Lagrangian flatness
factors of $V^{L}_i$ also observed in KS \cite{KS} are consistent with
the observation that fluid element pairs travel close to each other
for long stretches of time and separate in sudden bursts
\cite{tabeling,KS}. Fluid element accelerations ${\bf a} \equiv
{D\over Dt} {\bf u}$ (where ${D\over Dt} \equiv {\partial\over
\partial t} + {\bf u}\cdot {\bf \nabla}$) are such that ${\bf \nabla}
\cdot {\bf a} = {\bf s}^2 - {{\bf \omega}^2 \over 2}$ where ${\bf s}$
is the strain rate matrix and ${\bf \omega}$ the vorticity
vector. Hence, ${\bf \nabla} \cdot {\bf a}$ is large and positive most
often in straining regions around hyperbolic points of the flow where
${\bf s}^2$ is large and ${\bf \omega}^2$ close to 0. Close fluid
element pairs can separate violently where ${\bf \nabla} \cdot {\bf
a}$ is large and positive, and the separation is effective if the
streamline structure of the turbulence is persistent enough in
time. Hence, such violent separation events will most often occur where close
fluid element pairs meet hyperbolic points that are persistent
enough. 

Based on their KS results which were limited to $\Delta_0 = \eta/2$,
Fung \& Vassilicos (1998) \cite{KS} rephrased Richardson's locality
assumption as follows: ``in the inertial range, the dominant
contribution to the turbulent diffusivity ${d\over dt}\overline{
\Delta^{2}}$ comes from straining regions of size $\sqrt{\overline{
\Delta^{2}}}$; these straining regions are embedded in a fractal-eddy
structure of cat's eyes within cat's eyes and therefore straining
regions exist with a variety of length-scales over the entire inertial
range.'' Davila \& Vassilicos \cite{davilicos} have related $\gamma$
to the fractal dimension $D$ of this fractal-eddy streamline structure
of straining regions when $\Delta_0 $ is close to and below $\eta$
($\gamma = 4/D$). These results suggest that when $\Delta_0$ is
between $\eta$ and $0.1\eta$, the evolution of fluid element pairs by
bursts when they meet straining regions somehow tunes into the
straining fractal structure of the flow and gives rise to Richardson's
law. This requires some persistence of the streamline structure, and
indeed Richardson's law is lost when the unsteadiness parameter
$\lambda$ is made significantly larger than 1 \cite{KS}.

This topological picture of turbulent pair diffusion suggested by
results in previous papers and our argument concerning ${\bf \nabla}
\cdot {\bf a}$ could also explain the strong $\Delta_0$ - dependence
of $\overline{ \Delta^{2}}$. As $\Delta_0$ decreases well below
$\eta$, the probability for fluid element pairs to encounter a
hyperbolic point and be separated by it also decreases and can become
so small for $\Delta_0 \ll \eta$ that pairs may travel close to each
other for very long times. Eventually, at times nearing $L/u'$, the
eddy turnover time of the turbulence, the two fluid elements will be
separated by the unsteadiness of the flow rather than by its
streamline structure as they will have to become independent at times
$t\gg L/u'$. They therefore largely bypass the relatively persistent
straining fractal streamline structure of the turbulence and also
Richardson's law as a result. For initial conditions $\Delta_0 \gg
\eta$, the argument based on ${\bf \nabla} \cdot {\bf a}$ does not
apply and the separation of fluid element pairs cannot be considered
to be dominated by straining events in the vicinity of hyperbolic
regions.  In the framework of the topological turbulent pair diffusion
picture, this is consistent with the absence of a Richardson law for
$\Delta_0 \gg \eta$.

F.N. and J.C.V. are grateful for financial support from the Royal
Society and EPSRC.


\begin{thebibliography}{0}


\bibitem{ottino} J.M. Ottino, 
{\it The kinematics of mixing: stretching, chaos and transport} (Cambridge 
University Press, Cambridge, 1989). 

\bibitem{richardson} L.F. Richardson, Proc. R. Soc. London Sect. A {\bf 110},
709 (1926).


\bibitem{batchelor} G.K. Batchelor, Proc. Cambridge Philos. Soc. {\bf
48}, 345 (1952).

\bibitem{kraichnan} R.H. Kraichnan, Phys. Fluids {\bf 9}, 1937 (1966).

\bibitem{shlesinger} M.F. Shlesinger, B.J. West and J. Klafter, Phys. Rev. 
Lett. {\bf 58}, 1100 (1987).

\bibitem{tabeling} M.-C. Jullien, J. Paret and P. Tabeling, Phys. Rev. Lett. 
{\bf 82}, 2872 (1999).

\bibitem{KS} J.C.H. Fung {\it et al.}, J. Fluid Mech. {\bf 236} 281
(1992); J.C.H. Fung and J.C. Vassilicos, Phys. Rev. E {\bf 57} 1677
(1998); N.A. Malik and J.C. Vassilicos, Phys. Fluids {\bf 6} 1572
(1999).

\bibitem{fv00} P. Flohr and J.C. Vassilicos, J. Fluid Mech. (2000).
J. Fluid Mech. {\bf 407} 315 (2000).

\bibitem{romans} G. Boffetta, A. Celani, A. Crisanti and A. Vulpiani, Phys. 
Rev. E {\bf 60} 6734 (1999). 

\bibitem{sokolov} I.M. Sokolov, J. Klafter and A. Blumen, Phys. Rev. E 
{\bf 61} 2717 (2000).  

\bibitem{boffetta} G. Boffetta, In ``Advances In Turbulence VIII'' (ed. C. 
Dopazo {\it et al.}), CIMNE, Barcelona (2000). 

\bibitem{davilicos} J. Davila \& J.C. Vassilicos, preprint to be submitted, 2002.

\end{thebibliography}
\end{document}